\documentclass[10pt]{article}	
\usepackage{amsmath}
\usepackage{amsfonts}
\usepackage[top=1in,bottom=1in,left=1in,right=1in]{geometry}

\author{Yunsheng Li \and Chi Jin}
\date{\today}
\title{An Extended Small-Gain Theorem}

\begin{document}
\bibliographystyle{plain}
\maketitle
\newtheorem{ass}{Assumption}[section]
\newtheorem{rmk}{Remark}
\newtheorem{claim}{Claim}[section]
\newtheorem{dfn}{Definition}[section]
\newtheorem{theorem}{Theorem}[section]

\section{Introduction}
	In \cite{JJTTPP} a interconnected system consisted of two sub-systems is studied, which leads to a generalized small-gain theorem, which gives the express of the gain of the output of the interconnected system w.r.t external input signals. However, sometimes we are interested in the gain from the input to the output of each sub-system. Such a result will offer more flexibility in control synthesis. For this purpose, in this article we provide a modified version of the Theorem 2.1 in \cite{JJTTPP}.
	\subsection*{Facts and Notations}
	\begin{itemize}
		\item $|\cdot|$ stands for any vector norm.
		\item $Id$ denotes the identity function.
		\item $||u||$ denotes the $ess.sup.\{|u(t)|,t\geq0\}$ for any measurable function $u:\mathbb{R}_+\rightarrow\mathbb{R}^m$.
		\item $u_{[t_1,t_2]}$ denotes the truncation of the function $u$:
			\begin{equation}
			y=\left\{
				\begin{array}{ll}
				u(t) &\textrm{if $t\in[t_1,t_2]$}\\
				0 &\textrm{otherwise}\\
				\end{array}
			\right.
			\end{equation}
			and $u_T=u_{[0,T]}$.
		\item \textbf{Weak triangular inequality:} 
			\begin{equation}
			\gamma(a+b)\leq\gamma\circ(Id+\rho)(a)+\gamma\circ(Id+\rho^{-1})(b)\label{weaktrieq}
			\end{equation}
			$\gamma$ and $\rho$ are functions of class $K$ and $K_{\infty}$, respectively. $a$ and $b$ are non-negative real numbers. In particular, let $\rho=Id$, we get: $\gamma(a+b)\leq\gamma(2a)+\gamma(2b)$.
		\item $\rho$ is any function of class $K_{\infty}$, then:
			\begin{equation}
			[Id-(Id+\rho)^{-1}]^{-1}=Id+\rho^{-1}\label{invfun}
			\end{equation}
	\end{itemize}
\section{Definition and Main Results}
	\subsection*{Input-to-Output Practical Stability}
		Considering the following control system with $x$ as state, $u$ as input, and $y$ as output:
		\begin{equation}
		\label{defiops}
		\left\{
			\begin{array}{ll}
			\dot{x}=f(x,u) &\textrm{$x\in\mathbb{R}^n,u\in\mathbb{R}^m$}\\
			y=h(x,u) &\textrm{$y\in\mathbb{R}^p$}\\
			\end{array}
		\right.
		\end{equation}
		$f$ and $g$ are smooth functions.
		\begin{dfn}
		System (\ref{defiops}) is said to have the boundedness observability($UO$) property if a function $\alpha^0$ of class $K$ and a non-negative constant $D^0$ exist such that, for each measurable essentially bounded control $u(t)$ on $[0,T)$ with $0<T\leq+\infty$, the solution $x(t)$ of (\ref{defiops}) right maximally defined on $[0,T')$ satisfies:
			\begin{equation}
			\label{defuo}
			\left|x(t)\right|\leq\alpha^0(|x(0)|+\left\|(u_t^T,y_t^T)^T\right\|)+D^0, \forall t\in [0,T')
			\end{equation}
		\end{dfn}
		\begin{dfn}
		System (\ref{defiops}) is \textit{input-to-output practically stable(IOpS)} if a function $\beta$ of class $KL$, a function $\gamma$ of class $K$, called a \textit{(nonlinear) gain from input to output}, and a non-negative constant $d$ exist such that, for each initial condition $x(0)$, each measurable essentially bounded control $u(\cdot)$ on $[0,\infty)$ and each $t$ in the right maximal interval of definition of the corresponding solution of (\ref{defiops}), we have:
			\begin{equation}
			\left|y(t)\right|\leq\beta(|x(0)|,t)+\gamma(||u||)+d\label{iops}
			\end{equation}
		when (\ref{iops}) is satisfied with $d=0$, system (\ref{defiops}) is said to be \textit{input-to-output stable(IOS)}.
		\end{dfn}
	\subsection*{Generalized Small-Gain Theorem}
		Considering	now the following general interconnected system:
		\begin{equation}
			\label{sys}
			\begin{array}{cc}
				\dot{x_1}=f_1(x_1,y_2,u_1), & y_1=h_1(x_1,y_2,u_1)\\
				\dot{x_2}=f_2(x_2,y_1,u_2), & y_2=h_2(x_2,y_1,u_2)\\			
			\end{array}
		\end{equation}
		where, for $i=1,2,x_i\in\mathbb{R}^{n_i},u_i\in\mathbb{R}^{m_i}$ and $y_i\in\mathbb{R}^{p_i}$. The function $f_1,f_2,h_1$,and $h_2$ are smooth and a smooth function $h$ exists such that:
		\begin{equation}
			(y_1,y_2)=h(x_1,x_2,u_1,u_2)
		\end{equation}
		is the solution of 
		\begin{equation}
			\left\{
			\begin{array}{c}
			y_1=h_1(x_1,h_2(x_2,y_1,u_2),u_1),\\
			y_2=h_2(x_2,h_1(x_1,y_2,u_1),u_2).\\
			\end{array}
			\right.
		\end{equation}
		\begin{theorem}
		Suppose (\ref{sys}) is IOpS with $(y_2, u_1)$ $(resp. (y_1, u_2))$ as input, $y_1$ $(resp. y_2)$ as output, and\\
		$( \beta_1, ( \gamma_1^y, \gamma_1^u), d_1 )$ $(resp. (\beta_2, ( \gamma_2^y, \gamma_2^u ), d_2 ) )$ as triple satisfying (\ref{iops}), namely:
		\begin{equation}
		\begin{array}{ll}
			|y_1(t)|\leq\beta_1(|x_1(0)|,t)+\gamma_1^y(||y_{2t}||)+\gamma_1^u(||u_1||)+d_1\\
			|y_2(t)|\leq\beta_2(|x_2(0)|,t)+\gamma_2^y(||y_{1t}||)+\gamma_2^u(||u_2||)+d_2\\	
		\end{array}
		\end{equation}
		Also, suppose that (\ref{sys}) has the $UO$ property with couple $(\alpha_1^0, D_1^0)$ $(resp. (\alpha_2^0, D_2^0))$. If two functions $\rho_1$ and $\rho_2$ of class $K_\infty$ and a non-negative real number $s_l$ satisfying:
		\begin{equation}
			\label{conditions}
			\begin{array}{ll}
				\left. \begin{array}{c}
					(Id+\rho_2)\circ\gamma_2^y\circ(Id+\rho_1)\circ\gamma_1^y(s)\leq s,\\
					(Id+\rho_1)\circ\gamma_1^y\circ(Id+\rho_2)\circ\gamma_2^y(s)\leq s,\\
				\end{array}\right\}
			& \forall s\geq s_l
			\end{array}
		\end{equation}
		exist, then system (\ref{sys}) with $u=(u_1,u_2)$ as input, $y=(y_1,y_2)$ as output, and $x=(x_1,x_2)$ as state is $IOpS$ and has the $UO$ property (is $IOS$ and has the $UO$ property with $D^0=0$ when $s_l=d_i=D_i^0(i=1,2)$).
		
		More specifically, for each pair of class $K_\infty$ functions $(\rho_3, r_3^1)$, functions $\beta_1'$ and $\beta_2'$ of class $KL$, a function $r_3^2$ of class $K$, and non-negative constants $d_i'$(equal to zero when $s_l=d_i=D_i^0=0(i=1,2)$) exist such that the system (\ref{sys}) is $IOpS$ with the triple $(\beta_1'+\beta_2', r_1+r_2+r_3^1+r_3^2, d_1'+d_2')$ where
		\begin{equation}\label{r1r2}
		\left\{
		\begin{array}{c}
			r_1(s)=(Id+\rho_1^{-1})\circ(Id+\rho_3)^2\circ[\gamma_1^u+\gamma_1^y\circ(Id+\rho_2^{-1}\circ(Id+\rho_3)^2\circ\gamma_2^u)](s)\\
			r_2(s)=(Id+\rho_2^{-1})\circ(Id+\rho_3)^2\circ[\gamma_2^u+\gamma_2^y\circ(Id+\rho_1^{-1}\circ(Id+\rho_3)^2\circ\gamma_1^u)](s)\\
		\end{array}	
		\right.		
		\end{equation}				
		and the (nonlinear) gains from input $u$ to the output components $y_1(t)$ and $y_2(t)$ are $(r_1+r_3^1)$ and $(r_2+r_3^2)$, respectively.		
		\end{theorem}
		
		\begin{rmk}
		Furthermore, if the sub-systems have Input-to-State Stable property, namely:
		\begin{align}
			\left|x_1(t)\right| \leq \beta_{x1}(|x_1(0)|,t)+\gamma_{x1}^y(||y_{2t}||)+\gamma_{x1}^u(||u_1||)\\
			\left|x_2(t)\right| \leq \beta_{x2}(|x_2(0)|,t)+\gamma_{x2}^y(||y_{1t}||)+\gamma_{x2}^u(||u_2||)
		\end{align}
		where $\beta_{x1}$ and $\beta_{x2}$ are $KL$ functions, and $\gamma_{x1}^y$, $\gamma_{x1}^u$, $\gamma_{x2}^y$, $\gamma_{x2}^u$ are $K$ functions.
		And the $IOpS$ property is restricted to $IOS$ property, then we can deduce that the interconnected system is also $ISS$ and $IOS$.
		\end{rmk}
\section{Proof of Theorem}
	A first fact to be noticed is that (\ref{conditions}) implies the existence of a non-negative real number $d_3$ such that:
	\begin{equation}
	\label{first_fact}
		\begin{array}{cc}
			\left.
			\begin{array}{c}
			\gamma_2^y\circ(Id+\rho_1)\circ\gamma_1^y(s)\leq(Id+\rho_2)^{-1}(s)+d_3\\
			\gamma_1^y\circ(Id+\rho_2)\circ\gamma_2^y(s)\leq(Id+\rho_1)^{-1}(s)+d_3\\
			\end{array}
			\right\}
			& \forall s \geq 0
		\end{array}
	\end{equation}
	with $d_3=0$ when $s_l=0$.
	
	$Step\ 1: Existence \ and \ Boundedness \ of \ Solutions \ on\  [0,\infty)$. For any pair of measurable essentially bounded controls $(u_1(t),u_2(t))$ defined on $[0,\infty)$, for any initial condition $x(0)$, by hypothesis of smoothness, a unique solution $x(t)$ of (\ref{sys}) right maximally defined on $[0,T)$ with $T>0$ possibly infinite exists. Also, since (\ref{sys}) are $IOpS$, for any $\tau\in [0,T)$ and any
	\begin{equation}
		0\leq t_{10}\leq t_{20}\leq t_{11}\leq t_{21}<T-\tau
	\end{equation}
	we have, using time invariance and causality,
	\begin{eqnarray}
		\label{y1}
		\left|y_1(t_{11}+\tau)\right|\leq\beta_1(\left|x_1(t_{10}+\tau)\right|,t_{11}-t_{10})+\gamma_1^y(\left\|y_{2[t_{10}+\tau,t_{11}+\tau]}\right\|)+\gamma_1^u(\left\|u_1\right\|)+d_1\\
		\left|y_1(t_{21}+\tau)\right|\leq\beta_2(\left|x_2(t_{20}+\tau)\right|,t_{21}-t_{20})+\gamma_2^y(\left\|y_{1[t_{20}+\tau,t_{21}+\tau]}\right\|)+\gamma_2^u(\left\|u_2\right\|)+d_2\label{y2}
	\end{eqnarray}
	For ease of notation, set $\gamma_i=\gamma_i^y$ and $v_i=\gamma_i^u(\left\|u_i\right\|)$. Then pick an arbitrary $T_0\in[0,T)$ and let
	\begin{equation}
		\tau=t_{10}=t_{20}=0, \ t_{21}=T_0, \ t_{11}\in[0, T_0].
	\end{equation}
	By applying (\ref{weaktrieq}), (\ref{invfun}) and using (\ref{first_fact}), we get successively
		\begin{align}
		\left\|y_{2T_0}\right\| &\leq \beta_2(\left|x_2(0)\right|,0)+\gamma_2(\beta_1(\left|x_1(0)\right|,0)+\gamma_1(\left\|y_{2T_0}\right\|)+v_1+d_1)+v_2+d_2\\
		\begin{split}
		\  &\leq \beta_2(\left|x_2(0)\right|,0)+\gamma_2\circ(Id+\rho_1)\circ\gamma_1(\left\|y_{2T_0}\right\|)\\
		\  &\ \ \ +\gamma_2\circ(Id+\rho_1^{-1})(\beta_1(\left|x_1(0)\right|,0)+v_1+d_1)+v_2+d_2
		\end{split}\\
		\begin{split}
		\  &\leq \beta_2(\left|x_2(0)\right|,0)+(Id+\rho_2)^{-1}(\left\|y_{2T_0}\right\|)+d_3\\
		\  &\ \ \ +\gamma_2\circ(Id+\rho_1^{-1})(\beta_1(\left|x_1(0)\right|,0)+v_1+d_1)+v_2+d_2
		\end{split}\\
		\begin{split}\label{bound}
		\  &\leq (Id+\rho_2^{-1})(\beta_2(\left|x_2(0)\right|,0)+d_3\\
		\  &\ \ \ +\gamma_2\circ(Id+\rho_1^{-1})(\beta_1(\left|x_1(0)\right|,0)+v_1+d_1)+v_2+d_2)
		\end{split}
		\end{align}
	Since $T_0$ is arbitrary in $[0,T)$ and the right-hand side of (\ref{bound}) is independent of $T_0$, $y_2(t)$ is bounded on $[0,T)$. By symmetry, the same argument shows that $y_1(t)$ is bounded on $[0,T)$. Since the $x_1-subsystem$ and $x_2-subsystem$ satisfy the $UO$ property, we conclude that $x_1(t)$ and $x_2(t)$ are bounded on $[0,T)$. It follows by contradiction that $T=+\infty$.
	
	$Step\ 2: The\ IOpS\ Property$. Continuing from (\ref{bound}), we can establish bounds on the outputs in the following manner. From (\ref{weaktrieq}), for any function $\rho_3$ of class $K_{\infty}$, we have
	\begin{align}
	\begin{split}
		\left|y_2(t)\right| &\leq (Id+\rho_2^{-1})(\beta_2(|x_2(0)|,0)+d_3+\gamma_2\circ(Id+\rho_1^{-1})\circ(Id+\rho_3^{-1})(\beta_1(|x_1(0)|,0))\\
		\  &\ \ \ +\gamma_2\circ(Id+\rho_1^{-1})\circ(Id+\rho_3)(v_1+d_1)+v_2+d_2)
	\end{split}\\
	\begin{split}
		\  &\leq (Id+\rho_2^{-1})\circ(Id+\rho_3^{-1})(\beta_2(|x_2(0)|,0)+\gamma_2\circ(Id+\rho_1^{-1})\circ(Id+\rho_3^{-1})(\beta_1(|x_1(0)|,0)))\\
		\  &\ \ \ +(Id+\rho_3)(d_3+\gamma_2\circ(Id+\rho_1^{-1})\circ(Id+\rho_3)(v_1+d_1)+v_2+d_2)
	\end{split}
	\end{align}
	So, by symmetry, we have established:
	\begin{equation}
	\label{bounds_on_outputs}
		\begin{array}{cc}
		\left|y_1(t)\right|\leq\delta_1(\left|x(0)\right|)+\Delta_1, &\left|y_2(t)\right|\leq\delta_2(\left|x(0)\right|)+\Delta_2
		\end{array}
	\end{equation}
	with			
	\begin{align}
	\label{delta}
		\begin{split}
		\delta_1(s)&=(Id+\rho_1^{-1})\circ(Id+\rho_3^{-1})(\beta_1(s,0)+\gamma_1\circ(Id+\rho_2^{-1})\circ(Id+\rho_3^{-1})(\beta_2(s,0)))\\
		\delta_2(s)&=(Id+\rho_2^{-1})\circ(Id+\rho_3^{-1})(\beta_2(s,0)+\gamma_2\circ(Id+\rho_2^{-1})\circ(Id+\rho_3^{-1})(\beta_2(s,0)))\\
		\Delta_1&=(Id+\rho_1^{-1})\circ(Id+\rho_3)(d_3+\gamma_1\circ(Id+\rho_2^{-1})\circ(Id+\rho_3)(v_2+d_2)+v_1+d_1)\\
		\Delta_2&=(Id+\rho_2^{-1})\circ(Id+\rho_3)(d_3+\gamma_2\circ(Id+\rho_2^{-1})\circ(Id+\rho_3)(v_1+d_1)+v_2+d_2)
		\end{split}
	\end{align}		
	With these bounds on the outputs we can use the $UO$ property to establish bounds on the states $x_i$. In particular, let $(\alpha_1^0,D_i^0),i=1,2$, be two couples satisfying (\ref{defuo}) respectively for the subsystems (\ref{sys}). In this case any solution $x(t)$ of (\ref{sys}) satisfies, for all $t\geq 0$:
	\begin{eqnarray}
		\label{bound_on_x1}
		\left|x_1(t)\right|\leq\alpha_1^0\left(\left|x_1(0)\right|+\left\|(u_{1t}^T,y_{2t}^T,y_{1t}^T)^T\right\|\right)+D_1^0\\
		\left|x_2(t)\right|\leq\alpha_2^0\left(\left|x_2(0)\right|+\left\|(u_{2t}^T,y_{1t}^T,y_{2t}^T)^T\right\|\right)+D_2^0\label{bound_on_x2}
	\end{eqnarray}
	From (\ref{bounds_on_outputs}), (\ref{bound_on_x1}), (\ref{bound_on_x2}), and (\ref{weaktrieq}), we have:
	\begin{align}
		\begin{split}\label{bound_on_x}
		\left\|x\right\| &\leq (\alpha_1^0+\alpha_2^0)\circ(2Id+2\delta_1+2\delta_2)(\left|x(0)\right|)+[(\alpha_1^0+\alpha_2^0)(2\left|u\right|+2\Delta_1+2\Delta_2)+D_1^0+D_2^0]\\
		\ &:=\delta_3(\left|x(0)\right|)+\Delta_3\\
		\ &:=s_{\infty}
		\end{split}
	\end{align}
	with $\delta_i$ and $\Delta_i(i=1,2)$ defined in (\ref{delta}).
	
	With these bounds on the states, inequalities (\ref{bounds_on_outputs}) can be completed as follows: Let
	\begin{equation}
	 \begin{array}{cccc}
	 t_{10}=\displaystyle{\frac{t}{4}}, &t_{20}=\displaystyle{\frac{t}{2}}, &t_{21}=t, &t_{11}\in[\displaystyle{\frac{t}{2}},t]
	 \end{array}
	\end{equation}
	and substitute (\ref{y1}) in (\ref{y2}), so that we have, for any $t \geq 0$ and $\tau \geq 0$,
	\begin{equation}
		\left|y_2(t+\tau)\right| \leq \beta_2\left(s_{\infty},\displaystyle{\frac{t}{2}}\right)+v_2+d_2+\gamma_2\left(\gamma_1\left(\left\|y_{2[t/4+\tau,\infty]}\right\|\right)+\beta_1\left(s_{\infty},\displaystyle{\frac{t}{4}}\right)+v_1+d_1\right)
	\end{equation}
	Thus, by applying (\ref{weaktrieq}) and using (\ref{first_fact}), we obtain, for all $t\geq 0$ and $\tan \geq 0$,
	\begin{align}
		\label{bracket}
		\begin{split}
		\left|y_2(t+\tau)\right| &\leq \left[\beta_2\left(s_{\infty},\displaystyle{\frac{t}{2}}\right)+\gamma_2\circ(Id+\rho_1^{-1})\circ(Id+\rho_3^{-1})\left(\beta_1\left(s_{\infty},\displaystyle{\frac{t}{4}}\right)\right)\right]\\
		\ &\ \ \ +(Id+\rho_2)^{-1}\left(\left\|y_{2[t/4+\tau,\infty]}\right\|\right)+\gamma_2\circ(Id+\rho_1^{-1})\circ(Id+\rho_3)(v_1+d_1)+v_2+d_2+d_3
		\end{split}
	\end{align}
	Note that the term between brackets in (\ref{bracket}) is a function if class $KL$ with respect to $(s_{\infty},t)$. Further,
	\begin{equation}
	\gamma_2\circ(Id+\rho_1^{-1})\circ(Id+\rho_3)(v_1+d_1)+v_2+d_2+d_3=[(Id+\rho_1^{-1})\circ(Id+\rho_3)]^{-1}(\Delta_2)
	\end{equation}
	So we apply $Lemma\ A.1$ of \cite{JJTTPP} to (\ref{bracket}) with $\tau$ fixed, $z(t)=|y_2(t+\tau)|, \mu=\displaystyle{\frac{1}{4}}, \lambda=(Id+\rho_3)$, and $\rho=(Id+\rho_2)^{-1}$. It follows, using symmetry, that two functions $\widehat{\beta}_1$ and $\widehat{\beta}_2$ of class $KL$ exist such that, for all $t \geq 0$ and $\tau \geq 0$,
	\begin{equation}
		\label{output_component}
		\begin{array}{cc}
		\left|y_1(t+\tau)\right| \leq \widehat{\beta}_1(s_{\infty},t)+\Delta_1,	&\left|y_2(t+\tau)\right| \leq \widehat{\beta}_2(s_{\infty},t)+\Delta_2\\
		\end{array}
	\end{equation}
	
	Now we define the following function on $\mathbb{R}_+^3$:
	\begin{equation}
		\sigma_1(s,\Delta,t):=min\{\widehat{\beta}_1(\delta_3(s)+\Delta,t),\delta_1(s)\}
	\end{equation}
	then, for any function $\alpha$ of class $K_{\infty}$ and for each $(s,\Delta,t)$, we have:
	\begin{align}
		\begin{split}\label{sigma1}
		\sigma_1(s,\Delta,t) &\leq \sigma_1(s,\alpha^{-1}(s),t)+\sigma_1(\alpha(\Delta),\Delta,t)\\
		\ &\leq \widehat{\beta}_1(\delta_3(s)+\alpha^{-1}(s),t)+\delta_1\circ\alpha(\Delta)
		\end{split}
	\end{align}
	In view of (\ref{bound_on_x}) and (\ref{sigma1}), (\ref{bounds_on_outputs}) and (\ref{output_component}) imply, for all $t\geq 0$,
	\begin{align}
		\label{y11}
		\begin{split}
		\left|y_1(t)\right| &\leq \sigma_1\left(|x(0)|,\Delta_3,t\right)+\Delta_1\\
		\ &\leq \widehat{\beta}_1\left(\delta_3(|x(0)|)+\alpha^{-1}(|x(0)|),t\right)+\delta_1\circ\alpha(\Delta_3)+\Delta_1
		\end{split}
	\end{align}
	Based on the notation (\ref{r1r2}) we can get:
	\begin{align}
		\label{delta1}
		\Delta_1 &\leq r_1\left(\left\|u\right\|\right)+\tilde{d}_1\\
		\label{delta2}
		\Delta_2 &\leq r_2\left(\left\|u\right\|\right)+\tilde{d}_2\\
		\Delta_3 &\leq \left(\alpha_1^0+\alpha_2^0\right)\circ(4Id+4r_1+4r_2)\left(\left\|u\right\|\right)+\left(\alpha_1^0+\alpha_2^0\right)(4\tilde{d}_1+4\tilde{d}_2)+D_1^0+D_2^0
	\end{align}
	where
	\begin{align}
		\tilde{d}_1 &=(Id+\rho_1^{-1})\circ(Id+\rho_3)\circ(Id+\rho_3^{-1})[d_1+d_3+\gamma_1^y\circ(Id+\rho_2^{-1})\circ(Id+\rho_3)\circ(Id+\rho_3^{-1})(d_2)]\\
		\tilde{d}_2 &=(Id+\rho_2^{-1})\circ(Id+\rho_3)\circ(Id+\rho_3^{-1})[d_2+d_3+\gamma_2^y\circ(Id+\rho_1^{-1})\circ(Id+\rho_3)\circ(Id+\rho_3^{-1})(d_1)]
	\end{align}
	Then, using (\ref{weaktrieq}) again, we have
	\begin{align}
	\begin{split}
	\delta_1\circ\alpha(\Delta_3) &\leq \delta_1\circ\alpha\circ(2\alpha_1^0+2\alpha_2^0)\circ(4Id+4r_1+4r_2)(\left\|u\right\|)\\
	\ &\ \ \ +\delta_1\circ\alpha((2\alpha_1^0+2\alpha_2^0)(4\tilde{d}_1+4\tilde{d}_2)+2D_1^0+2D_2^0)
	\end{split}
	\end{align}
	Now, giving any function $r_3^1$ of class $K_{\infty}$, we can chose $\alpha$ such that:
	\begin{equation}
		\label{delta33}
		\begin{array}{cc}
		\delta_1\circ\alpha\circ(2\alpha_1^0+2\alpha_2^0)\circ(4Id+4r_1+4r_2)(s)\leq r_3^1(s), &\forall s\geq 0
		\end{array}
	\end{equation}
	(for example, $\alpha=(Id+\delta_1)^{-1}\circ r_3^1\circ(Id+(2\alpha_1^0+2\alpha_2^0)\circ(4Id+4r_1+4r_2))^{-1}$).\\
	This in conjunction with (\ref{y11}), (\ref{delta1}), (\ref{delta2}) and (\ref{delta33})implies:
	\begin{equation}
	\left|y_1(t)\right| \leq \beta_1'(|x(0)|,t)+(r_1+r_3^1)(\left\|u\right\|)+d_1'
	\end{equation}
	where
	\begin{align}
		\beta_1'(s,t) &=\widehat{\beta}_1(\delta_3(s)+\alpha^{-1}(s),t)\\
		d_1' &=\tilde{d}_1+\delta_1\circ\alpha((2\alpha_1^0+2\alpha_2^0)(4\tilde{d}_1+4\tilde{d}_2)+2D_1^0+2D_2^0)
	\end{align}
	By symmetry, we have:
	\begin{equation}
	\left|y_2(t)\right| \leq \beta_2'(|x(0)|,t)+(r_2+r_3^2)(\left\|u\right\|)+d_2'
	\end{equation}
	where
	\begin{align}
		r_3^2 &=\delta_2\circ\delta_1^{-1}\circ r_3^1\\
		\beta_2'(s,t) &=\widehat{\beta}_2(\delta_3(s)+\alpha^{-1}(s),t)\\
		d_2' &=\tilde{d}_2+\delta_2\circ\alpha((2\alpha_1^0+2\alpha_2^0)(4\tilde{d}_1+4\tilde{d}_2)+2D_1^0+2D_2^0)
	\end{align}
	Finally, with the fact that $|y(t)|\leq|y_1(t)|+|y_2(t)|$ we obtain the $IOpS$ property for system (\ref{sys}) with the triple $(\beta_1'+\beta_2', r_1+r_2+r_3^1+r_3^2, d_1'+d_2')$.
	When $d_i=D_i^0=0(i=1,2)$ and $d_3=0(i.e., s_l=0)$, we get $d_1'=d_2'=0$ implying the $IOS$ property holds. The $UO$ property for the interconnection follows from the $UO$ property for each subsystem.

\bibliography{bibfile}
\end{document}